\documentclass{webofc}
\usepackage[varg]{txfonts}
\usepackage{hyperref}
\newcommand{\Sp}[1]{\ensuremath{\text{Sp}(#1)}}
\newcommand{\SU}[1]{\ensuremath{\text{SU}(#1)}}
\newcommand{\uone}{\ensuremath{\text{U}(1)}}
\newcommand{\uoneprime}{\ensuremath{\text{U}(1)'~}}
\graphicspath{ {figures/} }

\begin{document}
\title{Strongly Interacting Dark Matter from $\rm{Sp}(4)$ Gauge Theory}

\author{
    \firstname{Fabian}   \lastname{Zierler}  \inst{1} \fnsep\thanks{\email{fabian.zierler@uni-graz.at}}  \and
    \firstname{Suchita}  \lastname{Kulkarni} \inst{1} \and 
    \firstname{Axel}     \lastname{Maas}     \inst{1} \and 
    \firstname{Seán}     \lastname{Mee}      \inst{1} \and 
    \firstname{Marco}    \lastname{Nikolic}  \inst{2} \and 
    \firstname{Josef}    \lastname{Pradler}  \inst{2} 
}

\institute{
    Institute of Physics, NAWI Graz, University of Graz, Universit\"atsplatz 5, A-8010 Graz, Austria \and 
    Institute of High Energy Physics, Austrian Academy of Sciences, Nikolsdorfergasse 18, 1050 Vienna, Austria
}

\abstract{The stable hadronic bound states in a hidden new non-Abelian gauge sector provide interesting candidates for strongly-interacting Dark Matter (DM). A particular example are theories in which DM is made up of dark pions which set the DM relic abundance through self-annihilation. One of the simplest realizations is $\Sp{4}_c$ gauge theory with two Dirac fermions. We discuss its mesonic multiplets for degenerate and non-degenerate fermions, construct a low-energy effective theory and present lattice results for the pseudoscalar mesons and vector mesons.}

\maketitle
\section{Introduction}
\label{intro}Non-Abelian, confining, gauge theories with fermions can provide composite Dark Matter (DM) candidates. For a sufficient number of pseudo-Nambu-Goldstone bosons (pNGBs) $\pi$ of chiral symmetry breaking the relic density of Dark Matter in the current universe can be obtained through a $3  \to 2$ self-annihilation freeze-out process \cite{Hochberg:2014dra, Hochberg:2014kqa}. These models have been termed Strongly Interacting Massive Particles (SIMP) models. A minimal model of SIMP Dark Matter is given by $\Sp{4}_c$ gauge theory with $2$ fundamental, massive fermions. Due to the pseudo-reality of the fundamental representation of $\Sp{4}_c$  these theories have exactly $5$ pNGBs which is the minimum number needed for the SIMP mechanism. In the language of chiral perturbation theory, this interaction is provided through the Wess-Zumino-Witten (WZW) term.

In addition, a mediator to the Standard Model (SM) is required to thermally couple the dark sector to the SM.  Here, we study the case of a dark photon $Z'$ associated with an additional \uoneprime gauge theory as mediator. The dark fermions are charged under this interaction and the $Z'$ mixes kinetically with the SM model photon. All other $\Sp{N_c=2N}$ theories with $N_f=2$ provide the WZW term needed for SIMP DM. However, larger gauge groups than $\Sp{2}_c=\SU{2}_c$ appear to be phenomenologically favoured \cite{Hochberg:2014kqa}. Therefore, we study $\Sp{4}_c$ as the next-to-minimal case.

The stability of the pNGBs in isolation is provided through the global $\Sp{4}$ flavour symmetry after chiral symmetry breaking. Initial investigations suggest that this model is significantly less constrained if the global symmetry is further broken \cite{Hochberg:2014kqa}. We thus investigate two different sources for such a breaking of the global symmetry: 1) The charge assignments of the dark fermions under the dark \uoneprime and 2) strong isospin breaking in the dark sector through non-degenerate fermion masses. In both cases we discuss the remaining global symmetries and the multiplet structure of the mesons under them. Since the dark sector is strongly interacting and confining we need to treat the dark sector non-perturbatively. We make use of existing $\Sp{4}_c$ lattice results with two degenerate fermions \cite{Bennett:2019jzz} and extend them with our own dedicated simulations for non-degenerate masses. Using these results, we construct a consistent low-energy effective theory (EFT) for degenerate masses and non-degenerate masses, including the WZW term, the spin-1 mesons and pseudoscalar flavour-singlet meson $\eta'$. We extend the EFT in the degenerate case to include the dark \uoneprime and calculate the pNGB mass-splitting through radiative corrections. 
 
More details and further results can be found in \cite{Kulkarni:2022bvh,Zierler:2022qfq}.

\section{Global symmetries and pNGB multiplets}

In isolation the microscopic Lagrangian of \Sp{4} gauge theory with $N_f=2$ fundamental fermion fields is given by 
\begin{align}
    \mathcal L = -\frac{1}{2} \text{Tr} \left[ G_{\mu \nu} G^{\mu \nu} \right] + \bar u \left( \gamma_\mu D_\mu + m_u \right) u + \bar d \left( \gamma_\mu D_\mu + m_d \right) d.
\end{align}
We denote the fermion fields as $u$ and $d$ in analogy to QCD. Both, the gauge bosons and the dark fermions are singlets under any SM charge. The fundamental representation of symplectic gauge theories is pseudo-real. Hence, fermions and anti-fermions cannot be distinguished \cite{Kosower:1984aw}. For vanishing masses the global symmetry is given by $\SU{4}$. This symmetry is broken spontaneously by the fermion condensate and explicitly by degenerate masses $m_d = m_u$ down to $\Sp{4}$. If the fermions have distinct masses then we are left with a global $\SU{2} \times \SU{2}$ symmetry. 

Coupling the theory to an external \uoneprime vector field, $V^\mu$ the symmetry breaking depends on the charge assignment matrix $Q$ of the individual Weyl components of the fermions. On the level of the Lagrangian this is given by 
\begin{align}
   \mathcal L_{\rm break} = V^\mu \Psi^\dagger Q \partial_\mu \Psi, 
\end{align}
where $\Psi$ contains the left- and right-handed projections $q_{L/R}^T = (u_{L/R}^T, d_{L/R}^T)$ of the Dirac fermions in the form $\Psi^T = \begin{pmatrix} q_L^T , ( \sigma_2 S q_R )^T \end{pmatrix}$. Here, $\sigma_2$ is the second Pauli matrix that arises from expressing the gamma matrices in the chiral representation in a block diagonal from and $S$ is a colour matrix fulfilling $(\tau^a)^T =S \tau^a S$ for all generators $\tau^a$ of the gauge theory. This is always possible for pseudo-real representations. We find that all charge assignments $Q$ lead to either an unbroken $\SU{2} \times \uone $  subgroup of the global symmetry or an unbroken $\uone \times \uone$.

Due to the extended global symmetry the hadronic multiplets are enlarged by additional colour neutral quark-quark and antiquark-antiquark bound states (diquarks). No fermionic bound states exist and the low-energy degrees of freedom are mesonic. The collection of mesons, diquarks and antidiquarks appear either in $5$-plets, such as the pNGBs $\pi$, $10$-plets like the vector mesons $\rho$, or as singlets \cite{Bennett:2019jzz}. For our purposes the breaking pattern of our DM candidates -- the pNGBs $\pi$ -- are of special interest.
For non-degenerate masses the flavour diagonal pNGB $\pi^C$ is always a flavour-singlet while the other pNGBs form a $4$-plet -- see figure \ref{fig:kappa_multiplets} \cite{Kulkarni:2022bvh}.

For a charge assignment leading to a preserved $\uone \times \uone$ the $\pi^C$ is again a singlet and the other pNGBs form two doublets. Only degenerate fermions with charge assignments leading to a global $\SU{2}\times \uone$ result in a triplet and a doublet of pNGBs which guarantees stable DM candidates. In the other cases extra attention has to be paid to the lifetime of the non-singlet pNGB. It needs to be sufficiently long-lived to preserve the $3 \to 2$ interaction responsible for freeze-out. If they decay after freeze-out the relic density of DM is set by only the non-singlet pNGBs. \cite{Katz:2020ywn}  

\section{EFT construction for $\Sp{4}_c$ with $N_f=2$ and a dark photon portal}

The pNGBs become massless in the chiral limit and should remain the lightest states in the mesonic spectrum for non-vanishing moderate fermion masses. Depending on details of the spectrum, other states need to be included in the construction of an EFT. Initial lattice results on $\Sp{4}_c$ with $N_f=2$ suggests that the pNGBs and the vector mesons $\rho$ are the lightest non-singlet states \cite{Bennett:2019jzz} and studies of pure Yang-Mills theory indicate heavy glueballs \cite{Bennett:2020qtj, Bennett:2021mbw}. Even for heavy fermions they are significantly lighter than the scalar or axial-vector non-singlet mesons. Lattice results on the singlet pseudoscalar $\eta'$ suggest that this state is lighter than the vector mesons but heavier than the pNGBs for moderately heavy fermions \cite{Zierler:2022qfq}. The same hierarchy of light vector and pseudoscalar mesons and relatively heavy non-singlet axial-vector and scalar mesons has been observed in $\SU{2}_c$ with $N_f=2$ \cite{Arthur:2016dir,Arthur:2016ozw}. In two-flavour $\SU{3}_c$ gauge theory the pattern $m_\pi < m_{\eta'} \approx m_{\rho}$ has been observed in the chiral limit \cite{Dimopoulos:2018xkm}.

This leads to an EFT that includes the pNGBs $\pi$, the corresponding WZW term, the interactions with the $\uoneprime$ vector field $V_\mu$ (including the term responsible for the induced mass-splitting of the pNGBs) as well as the vector mesons $\rho$ and the pseudoscalar singlet $\eta'$. Schematically, we construct an EFT of the form 
\begin{align}
   \mathcal L_{\rm EFT} = \mathcal L_\pi  &+ \mathcal L _{\eta'} + \mathcal L_\rho + \mathcal L_\pi^{\rm WZW} + \mathcal L_{\eta'}^{\rm WZW}  + \mathcal L_\rho^{\rm WZW} + \mathcal L_{\rm meson~interactions} \nonumber  \\  &+ \mathcal L_V^{\rm WZW}  + \mathcal L_{\rho V} + \mathcal L_{\pi V} + \mathcal{L}^{\rm split}_{\pi V}. 
\end{align}
The first line arises purely from the strong dynamics. It includes the dynamics of the light mesons, the influence on the WZW interaction and the meson-meson interactions. The corresponding low energy constants (LECs) of the EFT can be determined from first-principles lattice calculations of the strong sector in isolation. At first order the involved LECs are the meson masses and their decay constants. The included spectrum is sketched in the left panel of figure \ref{fig:kappa_multiplets}.

\subsection{Strong dynamics}
For the pNGBs we utilize the standard construction through the field $\Sigma=e^{i \pi / f_\pi} \Sigma_0 e^{i \pi^T / f_\pi}$ where $\Sigma_0$ gives the orientation of the chiral condensate and the pNGB matrix is spanned by the five generators of the $\SU{4}/\Sp{4}$ coset. This leads to kinetic and mass terms
\begin{align}
   \mathcal L_\pi = \frac{f_\pi^2}{4} \rm{Tr} \left[ \partial_\mu \Sigma  \partial^\mu \Sigma^\dagger  \right] -  \frac{\mu^3}{2} \left( \rm{Tr}\left[ M \Sigma \right] + h.c. \right) 
\end{align}
where $\mu$ is related to the chiral condensate as $\langle \bar u u + \bar d d  \rangle = \mu^3 \Sigma_0$ and $M$ is the fermionic mass matrix in the basis of $\Psi$. The corresponding WZW term can be written as the integral over a five dimensional disc $Q_5$
\begin{align}
   \mathcal L_\pi^{\rm WZW} = \frac{-iN_c}{240 \pi^2} \int_{Q_5} \rm{Tr}\left[ (\Sigma^\dagger d \Sigma)^ 5 \right]. 
\end{align}
We introduce non-degenerate fermion masses $m_u \neq m_d$ by introducing an explicit splitting between the decay constants $f_{\pi_3} \neq f_{\pi_{1,2,4,5}}$ corresponding to diagonal pNGB and the others. We also include a splitting between the fermion condensates $\mu_u^3 \neq \mu_d^3$. This induces a change in the vacuum alignment of the chiral condensate $\mu^3 \Sigma_0^{\rm{deg}} \to \mu_u^3 \left( \Sigma_0^{\rm{deg}} + \Delta \Sigma_0 \right)$ where the induced difference $\Delta \Sigma_0$ is proportional to $(\mu_d^3 - \mu_u^3)/\mu^3_u$ which leads to a rescaling of $\mathcal L_\pi^{\rm{WZW}}$ in the presence of fermion mass splittings. 

The pseudoscalar flavour-singlet meson is associated with the $\uone_A$. We include its terms $\mathcal L_{\eta'}$ and $\mathcal L_{\eta'}^{\rm WZW}$ in the EFT by extending the coset space to $\rm{U}(4)/(\Sp{4} \times \uone_A)$ and arrive at $\Sigma = \exp( 2i \pi / f_{\pi}) \exp ( 2i \eta' T^0 / f_{\eta'}) \Sigma_0$ where $T^0 = \mathbb{I}_{4\times4} / \sqrt{8}$. For degenerate masses no contributions to the WZW occur through the inclusions of the $\eta'$ at first order. In the case of non-degenerate fermions both the diagonal pNGB $\pi^C$ and the $\eta'$ are flavour-singlets. Kinematic mixing occurs which modifies $\mathcal L_\pi^{\rm WZW}$ and leads to an extra term involving the $\eta'$.

For the spin-1 interactions -- both the vector mesons $\rho$ and the $U(1)'$ field $V_\mu$ -- we only considered the case of degenerate fermion masses. 

For the vector mesons we use \textit{hidden local symmetry} as our construction principle which introduces a local $\SU{4}$ symmetry and requires local invariance of the EFT leading to the 15 spin 1 states $\rho_\mu = \sum_{a=1}^{15} \rho^a_\mu T^a$. These 15 states consist of the 10-plet of vector mesons $\rho$ and the $5$-plet of axial-vector mesons $a_1$ which are known to be substantially heavier than the vector mesons from lattice investigations \cite{Arthur:2016dir,Bennett:2019jzz}.  We arrive at
\begin{align}
   \mathcal L_\rho = -\frac{1}{2} \rm{Tr}\left[ \rho_{\mu\nu} \rho^{\mu\nu} \right] + \frac{m_\rho^2}{2} \rm{Tr}\left[ \rho_{\mu} \rho^{\mu} \right], 
\end{align}
where $\rho_{\mu\nu} = \partial_\mu \rho_\nu -  \partial_\nu \rho_\mu - ig_\rho [\rho_\mu, \rho_\nu]$. We include the interactions with the pNGB by a construction analogous to the covariant derivative as $\partial_\mu \Sigma \to \partial_\mu \Sigma + i g_\rho \left( \rho_\mu \Sigma + \Sigma \rho_\mu^T\right)$ where $g_\rho$ is a LEC. This construction leads to non-diagonal kinetic terms for the axial vectors which ensures their higher masses $m_{a_1}^2/m_{\rho}^2 = \left(1-g_\rho^2 f_\pi^2 /(2m_\rho^2) \right)^{-1}$. We also include the effect on the WZW term which was already worked out for the general $\SU{2N_f}/\Sp{2N_f}$ coset \cite{Brauner:2018zwr}. 

\subsection{Dark photon portal}
\begin{figure}
    \begin{tabular}{p{0.4\textwidth} p{0.5\textwidth}}
        \vspace{0pt} \includegraphics[width=0.35\textwidth]{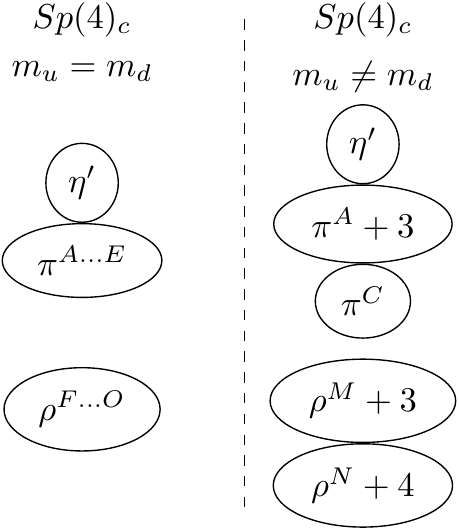} &
        \vspace{0pt} \includegraphics[width=0.49\textwidth]{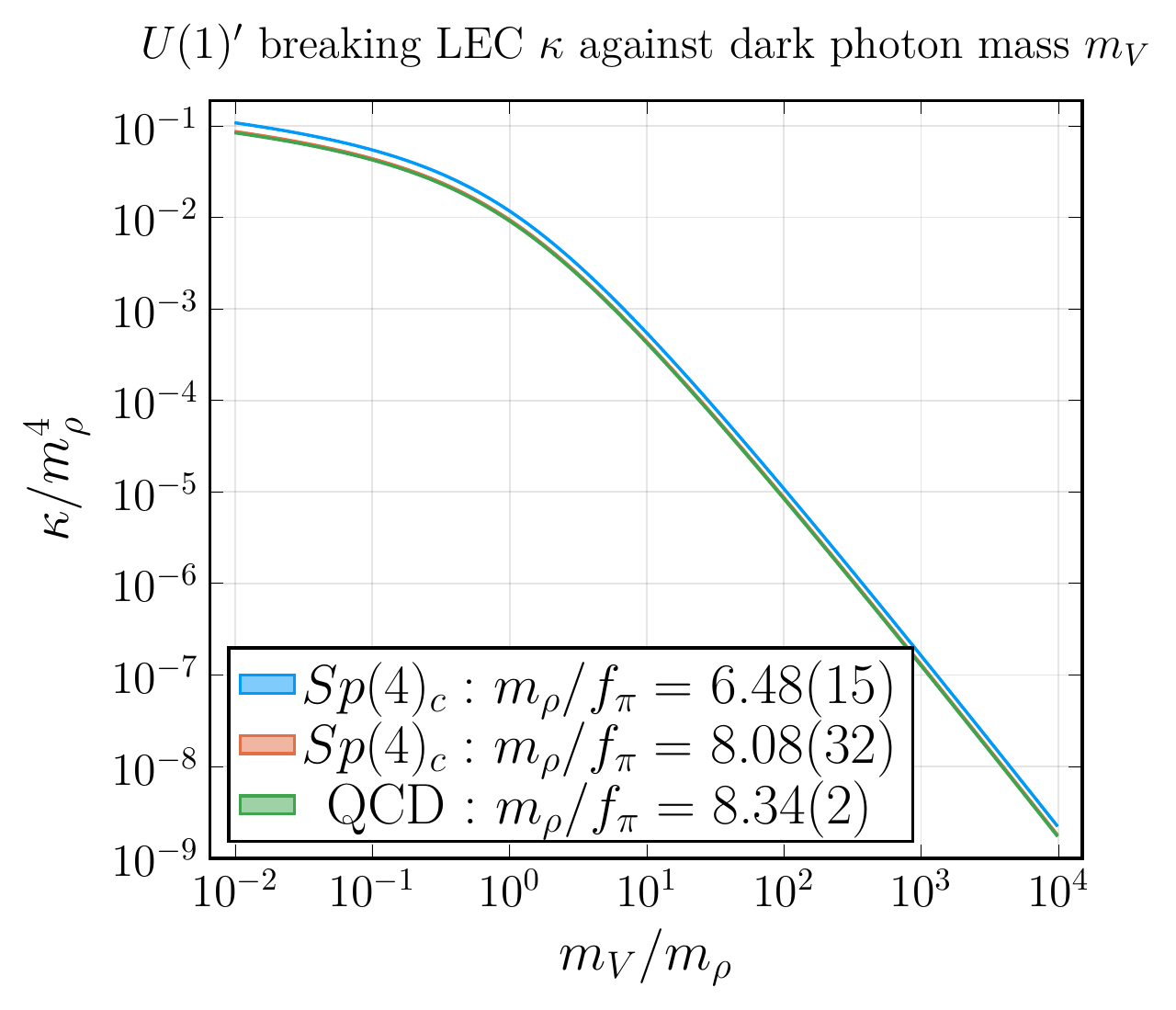}
    \end{tabular}
    \caption{(left) Meson multiplets of the pseudoscalar and vector mesons of the strong theory in isolation. The additional states compared to QCD are diquark states. (right) Constraints on the \uoneprime breaking parameter $\kappa$ from data taken from \cite{Bennett:2019jzz}. The value of $m_\rho / f_\pi = 6.48(1)$ corresponds to an ensemble with heavy fermions far away from the chiral limit and $m_\rho / f_\pi = 8.34(2)$ corresponds to the chiral extrapolation. We further provide the experimentally measured ratio of $m_\rho / f_\pi$ from SM QCD.}
    \label{fig:kappa_multiplets}
\end{figure}

The inclusion of the dark photon requires that we construct our EFT so that it is invariant under all local \uoneprime transformations. We restrict ourselves to diagonal charge assignment matrices $Q$. By examining the transformation properties of $\Sigma$ under \uoneprime we can determine the covariant derivative
\begin{align}
   D_\mu = \Sigma = \partial_\mu \Sigma + i e_D V_\mu \left( Q \Sigma + \Sigma Q^T\right) 
\end{align}
Depending on the exact charge assignment a subset of the pNGB couple to the dark photon and break the global $\Sp{4}$. Specifically, we can couple pairs of off-diagonal pNGBs. For one such assignment the $V_\mu - \pi$-interaction for the charged pNGBs $\pi^A$ and $\pi^B$ is given by
\begin{align}
   \mathcal L_{V\pi} = -2i e_D V^\mu \left( \pi^A \partial_\mu \pi^B - \pi^B \partial_\mu \pi^A \right) + e_D^2 V_\mu V^\mu \pi^A \pi^B 
\end{align}
and the interaction between the vector mesons and the dark photon is given by
\begin{align}
   \mathcal L_{\rho V} = -\frac{e_D}{g_\rho} V_{\mu\nu} \rm{Tr} \left[ Q \rho^{\mu \nu} \right] 
\end{align}
which preserves gauge invariance. Due to the presence of a global $U(1)$ symmetry after charge assignment the WZW term generically allows decays involving 3 particles such as $\pi \to V V$ \cite{Brauner:2018zwr}. However, the interaction 
\begin{align}
   \mathcal L_V^{\rm WZW} \supset \frac{40i C e_D^2}{f_\pi^2} \epsilon_{\mu  \nu \alpha \beta} V^\mu \partial^\nu V_\alpha \rm{Tr} \left[ Q^2 \partial^\beta \pi \right] 
\end{align}
is vanishing if the anomaly-cancellation condition $Q^2 \propto \mathbb I$ is implemented. Choosing the charge assignment such that all pNGBs transform non-trivially under the global symmetry group ensures that such interactions are vanishing to all orders. The induced mass splitting of the charged and uncharged Goldstones under the $U(1)'$ can be described by including the additional term 
\begin{align}
   \mathcal{L}^{\rm split}_{\pi V} = \kappa \rm{Tr}\left[Q \Sigma Q \Sigma^\dagger \right] 
\end{align}
which introduces a new LEC $\kappa$ parameterizing this effect. The splitting takes the form $\Delta m_\pi^2 = \kappa e_D^2 / f_\pi^2$. Assuming vector meson dominance yields
\begin{align}
    \Delta m_\pi^2 = \frac{6 e_D^2}{(2\pi)^2}\frac{m_\rho^4}{m_V^2 - m_\rho^2} \log \left( \frac{m_V^2}{m_\rho^2} \right).
\end{align}
This expression is exact in the chiral limit. It is independent of the hierarchy of the dark photon $V$ and the vector meson $\rho$ and continuos at $m_V = m_\rho$. Using this we can relate the LECs $\kappa, f_\pi$ and $m_\rho$. The latter two are not independent and their ratio is determined by the underlying UV complete theory. By using dimensionless ratio we can then constrain $\kappa / m_\rho^4$ as a function of $f_\pi / m_\rho$ from available lattice data from \cite{Bennett:2019jzz} as depicted in figure \ref{fig:kappa_multiplets}.

\section{Low Energy Constants from the Lattice}

The LECs cannot be determined from the EFT itself but must be obtained from the underlying, microscopic UV theory. For strongly-interacting, confining theories lattice field theory provides the tools to calculate the LECs from first principle. At leading order we are interested in the meson masses and their decay constant. For $\SU{2}_c$ and $\Sp{4}_c$ dedicated studies for two degenerate fermions have already been performed \cite{Arthur:2016dir,Arthur:2016ozw,Bennett:2019jzz,Zierler:2022qfq}. We supplement the results for $\Sp{4}_c$ with additional simulations at non-degenerate fermion masses.

We use the unimproved Wilson plaquette action and include two unimproved Wilson fermions using the Rational Hybrid Monte Carlo (RHMC) algorithm. We monitor the lowest eigenvalue of the fermion determinants and do not see any hints of a sign problem for the fermion masses studied here. At each lattice spacing we choose three different degenerate ensembles corresponding to different values of $m_\pi / m_\rho$. We then keep one fermion mass fixed and increase the other. We do this for three different inverse couplings $\beta$. The masses and decay are extracted from the large Euclidean $t$ behaviour of the meson correlator. We determine the renormalization factors for the decay constants using lattice perturbation theory at one loop as in \cite{Bennett:2019jzz}. Furthermore, we determine the scheme-dependent quark mass through the partially conserved axial current (PCAC).

\begin{figure}
    \centering
    \includegraphics[width=.49\textwidth]{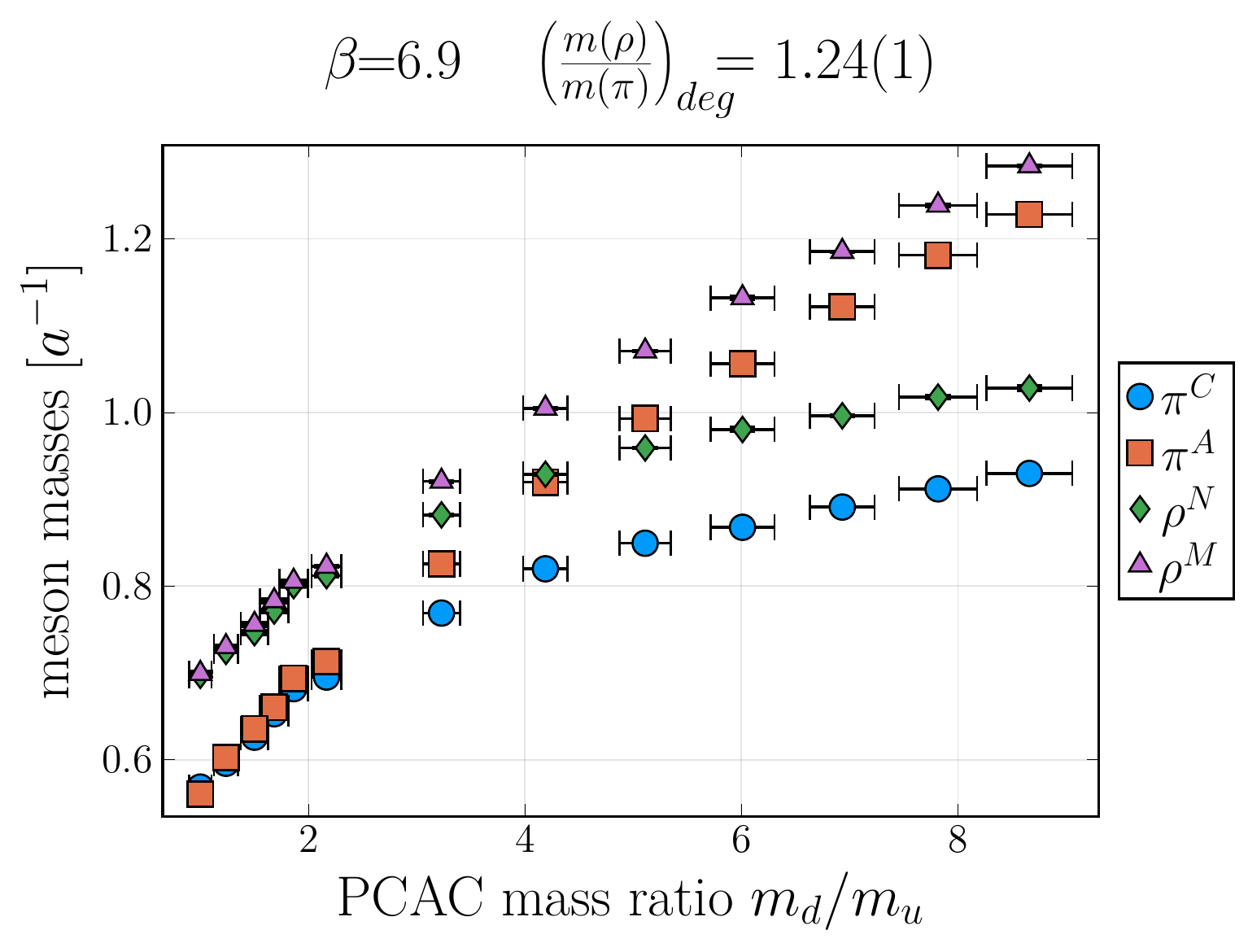}
    \includegraphics[width=.46\textwidth]{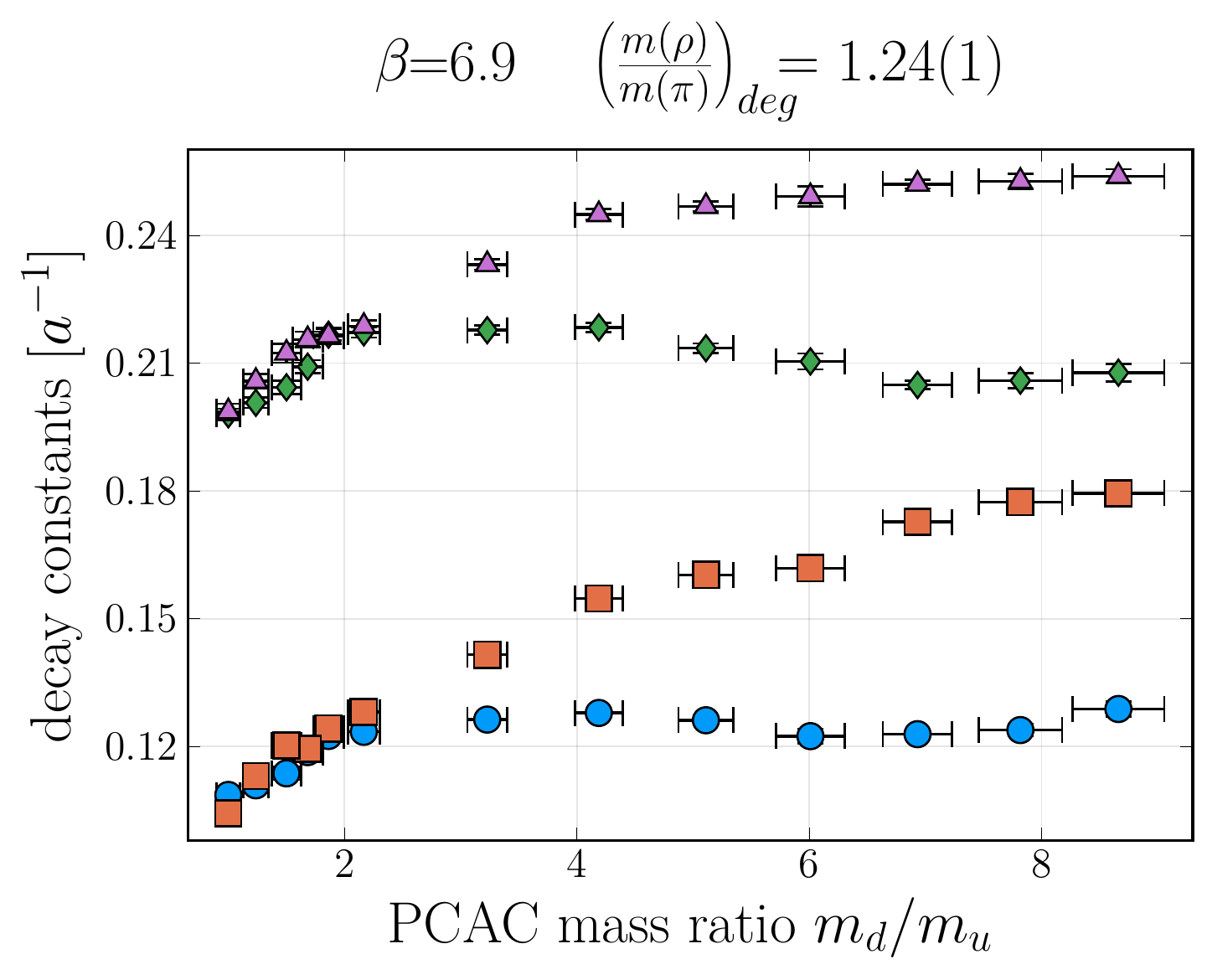}
    \caption{Masses and decay constants of the non-singlet pseudoscalar and vector mesons in lattice units. The unflavoured mesons are always lighter and the corresponding decay constants are smaller. This pattern holds for all other masses and inverse couplings $\beta$ studied so far.}
    \label{fig:lattice}
\end{figure}

We find that in all cases the unflavoured mesons to be lighter than their flavoured counterpart. The unflavoured pNGB $\pi^C$ is the lightest mesonic state of the theory. As we increase one fermion mass the unflavoured vector mesons get lighter than the flavoured pseudoscalars. Eventually, at large mass differences, the system resembles a heavy-light system. We show the results for on set of ensembles in figure \ref{fig:lattice}. We only see finite volume effects in the lightest ensembles on the finest lattices (i.e. at $\beta=7.2$ and $m^{\rm{bare}}_q \leq -0.78$) \cite{Kulkarni:2022bvh}. For the meson masses and the pseudoscalar decay constant we estimate our finite spacing effects to be smaller than $10\%$. For the vector decay constant they can be as large as $20\%$. Our results, together with those of \cite{Bennett:2019jzz}, suggest that the chiral effective theory should be applicable for $m_{\rho^C}/m_{\pi^C}>1.4$ and $m_d/m_u<1.5$ \cite{Kulkarni:2022bvh}.

\section{Summary}
We presented $\Sp{4}_c$ gauge theory with two fundamental fermions connected to SM via a \uoneprime dark photon as a candidate DM theory. We discussed the multiplet structure under a global $\Sp{4}$ and its breaking due to either non-degenerate fermions or \uoneprime charge assignments. We constructed an EFT for the strong theory in isolation including strong isospin breaking for the pNGBs $\pi$and the $\eta'$ meson. We further constructed an EFT for degenerate fermions including the vector mesons $\rho$ and the dark photon. We supplement existing lattice data for degenerate fermions by our own non-degenerate simulations. These results constrain the LECs in the EFTs presented here. 

\section*{Acknowledgements}
This work was supported by the FWF Austrian Science Fund research teams grant STRONG-DM (FG1). SK is supported by the FWF Austrian Science Fund Elise-Richter grant project number V592-N27. MN is supported by the Austrian Science Fund FWF under the Doctoral Program W1252-N27 Particles and Interactions. We thank E.~Bennett, H.~Kole\v{s}ov\'{a}, B.~Lucini, J.-W. Lee, M.~Piai and J.~Pomper for helpful discussions and the authors of \cite{Bennett:2017kga,Bennett:2019cxd,Bennett:2019jzz} for access to the $\Sp{2N}$ HiRep code prior to publication. The lattice results presented have been obtained using the Vienna Scientific Cluster (VSC4).

\bibliography{references}

\begin{thebibliography}{15}

\bibitem{Hochberg:2014dra}
Y.~Hochberg, E.~Kuflik, T.~Volansky, J.G. Wacker, Phys. Rev. Lett.
  \textbf{113}, 171301 (2014), \texttt{1402.5143}

\bibitem{Hochberg:2014kqa}
Y.~Hochberg, E.~Kuflik, H.~Murayama, T.~Volansky, J.G. Wacker, Phys. Rev. Lett.
  \textbf{115}, 021301 (2015), \texttt{1411.3727}

\bibitem{Bennett:2019jzz}
E.~Bennett, D.K. Hong, J.W. Lee, C.J.D. Lin, B.~Lucini, M.~Piai, D.~Vadacchino,
  JHEP \textbf{12}, 053 (2019), \texttt{1909.12662}

\bibitem{Kulkarni:2022bvh}
S.~Kulkarni, A.~Maas, S.~Mee, M.~Nikolic, J.~Pradler, F.~Zierler (2022),
  \texttt{2202.05191}

\bibitem{Zierler:2022qfq}
F.~Zierler, J.W. Lee, A.~Maas, F.~Pressler (2022), \texttt{2210.11187}

\bibitem{Kosower:1984aw}
D.A. Kosower, Phys. Lett. B \textbf{144}, 215 (1984)

\bibitem{Katz:2020ywn}
A.~Katz, E.~Salvioni, B.~Shakya, JHEP \textbf{10}, 049 (2020),
  \texttt{2006.15148}

\bibitem{Bennett:2020qtj}
E.~Bennett, J.~Holligan, D.K. Hong, J.W. Lee, C.J.D. Lin, B.~Lucini, M.~Piai,
  D.~Vadacchino, Phys. Rev. D \textbf{103}, 054509 (2021), \texttt{2010.15781}

\bibitem{Bennett:2021mbw}
E.~Bennett, J.~Holligan, D.K. Hong, H.~Hsiao, J.W. Lee, C.J.D. Lin, B.~Lucini,
  M.~Mesiti, M.~Piai, D.~Vadacchino, PoS \textbf{LATTICE2021}, 308 (2022),
  \texttt{2111.14544}

\bibitem{Arthur:2016dir}
R.~Arthur, V.~Drach, M.~Hansen, A.~Hietanen, C.~Pica, F.~Sannino, Phys. Rev. D
  \textbf{94}, 094507 (2016), \texttt{1602.06559}

\bibitem{Arthur:2016ozw}
R.~Arthur, V.~Drach, A.~Hietanen, C.~Pica, F.~Sannino (2016),
  \texttt{1607.06654}

\bibitem{Dimopoulos:2018xkm}
P.~Dimopoulos et~al., Phys. Rev. D \textbf{99}, 034511 (2019),
  \texttt{1812.08787}

\bibitem{Brauner:2018zwr}
T.~Brauner, H.~Kole\v{s}ov\'a, Nucl. Phys. B \textbf{945}, 114676 (2019),
  \texttt{1809.05310}

\bibitem{Bennett:2017kga}
E.~Bennett, D.K. Hong, J.W. Lee, C.J.D. Lin, B.~Lucini, M.~Piai, D.~Vadacchino,
  JHEP \textbf{03}, 185 (2018), \texttt{1712.04220}

\bibitem{Bennett:2019cxd}
E.~Bennett, D.K. Hong, J.W. Lee, C.J.D. Lin, B.~Lucini, M.~Mesiti, M.~Piai,
  J.~Rantaharju, D.~Vadacchino, Phys. Rev. D \textbf{101}, 074516 (2020),
  \texttt{1912.06505}

\end{thebibliography}
\end{document}